\documentclass[mathleft]{an}
\usepackage{graphicx,color}
\usepackage{times}
\newfont{\gwpfont}{cmssq8 scaled 1000}
\newcommand{\reflex}{{\gwpfont REFLEX}}
\newcommand{\noras}{{\gwpfont NORAS}}
\newcommand{\eros}{{\it eROSITA}}
\newcommand{\rosat}{{\it ROSAT}}

\begin{document}

% The following seven commands are intended for editorial usage and should be ignored by
% the author(s).
\Pagespan{789}{}% Document's page range. 
% If second parameter is left empty, the last page is computed automatically.
\Yearpublication{2012}%
\Yearsubmission{2012}%
\Month{08}%   
\Volume{999}%  
\Issue{88}% 
% \DOI{This.is/not.aDOI}% 

\title{Cluster science from {\it ROSAT} to {\it eROSITA}}

\author{G. Chon\inst{1}\fnmsep\thanks{Corresponding author:
  \email{gchon@mpe.mpg.de}\newline}
\and  H. B\"ohringer\inst{1}
}
\titlerunning{Cluster science from {\it ROSAT} to {\it eROSITA}}
\authorrunning{G. Chon \& H. B\"ohringer}
\institute{
Max-Planck-Institut f\"ur extraterrestrische Physik, 
85748 Garching, Germany
}

\received{24 August 2012}
\accepted{13 September 2012}
\publonline{later}


\abstract{
Galaxy clusters are one of the important cosmological probes to 
test the consistency of the observable structure and evolution of our Universe 
with the predictions of specific cosmological models. We use results 
from our analysis of the X-ray flux-limited \reflex\ cluster sample 
from the \rosat\ All-Sky Survey to illustrate the constraints on 
cosmological parameters that can be achieved with this approach. 
The upcoming \eros\ project of the Spektrum-Roentgen-Gamma mission 
will increase these capabilities by two orders of magnitude and 
importantly also increase the redshift range of such studies. 
We use the projected instrument performance to make predictions on 
the scope of the \eros\ survey and the potential of its exploitation.
}

\maketitle

\section{Introduction}

To perform cosmological tests with observations of the galaxy 
cluster population is one of the major goals of the \eros\ X-ray 
All-Sky Survey project (Predehl et al. 2011). Galaxy clusters offer 
themselves as one of a few important cosmological probes that can 
be used to constrain cosmological parameters essentially by assessing 
the statistics and growth of the large-scale structure in the Universe
(e.g. B\"ohringer et al. 2011, Vikhlinin et al. 2009). The detection 
of galaxy clusters in X-rays has the advantage that (i) the X-ray 
emitting plasma indicates a massive gravitationally bound object 
which can hold the hot plasma in place and (ii) the X-ray luminosity 
is tightly correlated to the cluster mass such that we have a method 
that is nearly mass selective. Therefore X-ray surveys are currently 
the most efficient approach for cosmological cluster studies.

The \rosat\ X-ray All-Sky Survey (RASS, Tr\"umper 1993) offers 
the best data base to construct a sample of the low redshift galaxy 
cluster population, which can be used for such cosmological tests. 
We therefore use our X-ray flux-limited RASS based galaxy cluster 
surveys \reflex\ and \noras\ here to illustrate the potential of 
galaxy cluster studies for cosmology. The \eros\ survey 
(Predehl et al. 2011) will have a sensitivity which is about 30 times 
better than that of the RASS and will not only dramatically increase 
the statistics of the cluster sample, but also bring in a new dimension 
by allowing us to study the evolution of the galaxy cluster population 
and the growth of cosmic structure over an extended redshift range up 
to $z \sim 1.5$. In the second part of this article we illustrate 
these capabilities of the German \eros\ instrument on board of the 
Russian Spektrum-Roentgen-Gamma space craft for cosmological and 
astrophysical galaxies cluster studies.

\section{The REFLEX galaxy cluster survey}

The \reflex\ cluster survey covering the southern sky below declination
+2.5 deg comprising currently 914 clusters with only 10 objects 
without spectroscopic redshifts is the best and most complete 
X-ray galaxy cluster sample available (Chon \& B\"ohringer 2012).
The \reflex\ II sample as a whole extends the flux limit of 
\reflex\ I from 3$\times10^{-12}$ erg/s/cm$^2$ to 1.8$\times10^{-12}$ 
erg/s/cm$^2$ in the 0.1-2.4 keV \rosat\ band. The construction
of the survey is described in detail by B\"ohringer
et al. (2012).

\begin{figure}
\begin{center}
\resizebox{\hsize}{!}{\includegraphics{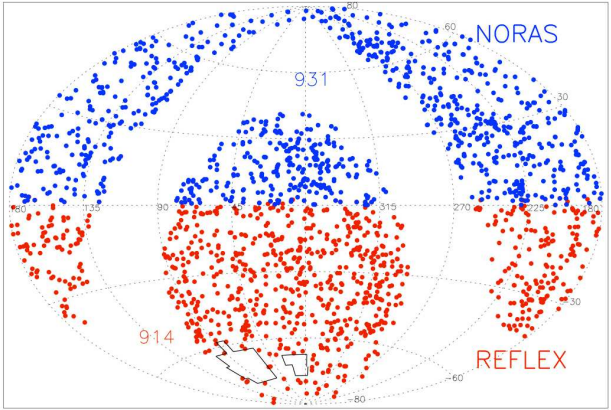}}
\end{center}
\caption{
Galaxy clusters identified in the \rosat\ All Sky Survey
within the \reflex\ II (red points) and \noras\ II
projects (blue points) with high completeness down
to a flux limit of 
$f_X>$1.8$\times 10^{-12}$ erg/s/$\mathrm {cm}^2$ in the 0.1-2.4 keV band.
The median redshift of the clusters is $z_{\mathrm med} \sim 0.1$
and the highest redshift is $z=0.54$. 
}
\label{fig:1}
\end{figure}

Fig. 1 shows the sky distribution of the extended \reflex\ survey 
(\reflex\ II) comprising 914 objects together with the northern
\noras\ sample. We have conducted a number of cosmological studies 
with the older \reflex\ I sample (B\"ohringer et al. 2001, 2004,
2011, Collins et al. 2001, Schuecker et al. 2003a,b) comprising
447 clusters. In the top panel of Fig. 2 we show the X-ray luminosity 
function for the clusters of \reflex\ I with a median redshift of 
$z = 0.08$ (B\"ohringer et al. 2002). We also show a fit of a 
cosmological model prediction for this function with the cosmological 
parameters of $\Omega_m = 0.28$, $\sigma_8 = 0.79$, a Hubble constant 
of $H_0 = 70$~km/s/Mpc and assuming a flat Universe, where the first 
two parameters have been varied for the fit. The mass function for the 
clusters has been calculated based on the recipe of Tinker et al. (2008) 
and converted to a luminosity function by means of the X-ray 
luminosity - mass relation from Reiprich and B\"ohringer (2002). 
In the lower panel we show how well the two cosmological parameters 
are constrained. These results provided the best constraints of the 
amplitude parameter of the matter density fluctuations in the Universe, 
$\sigma_8$, even before the first cosmic microwave results from the WMAP 
satellite became available. Combining the cluster abundance above a certain
X-ray luminosity with the large-scale spatial distribution of the \reflex\
clusters Schuecker et al. (2003a,b) obtained similar tight constraints and
showed in combination with data from distant supernovae (Perlmutter et al. 1999)
that the equation of state parameter, $w$, of Dark Energy is consistent with 
a value of -1 and thus also with a cosmological constant with an uncertainty 
of less than 30\%, a remarkable result when it was published in 2003. 

\begin{figure}
\begin{center}
\resizebox{\hsize}{!}{\includegraphics{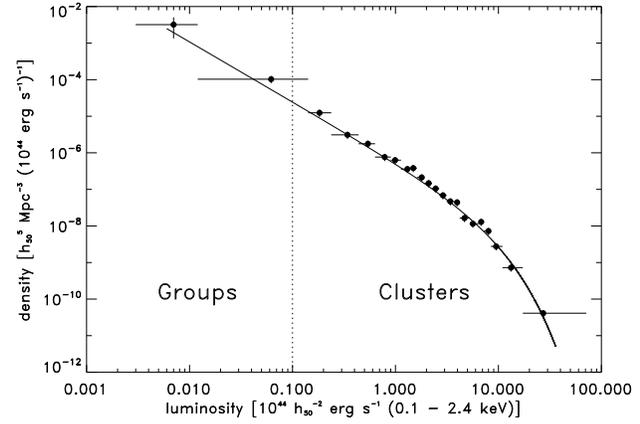}}
\resizebox{\hsize}{!}{\includegraphics{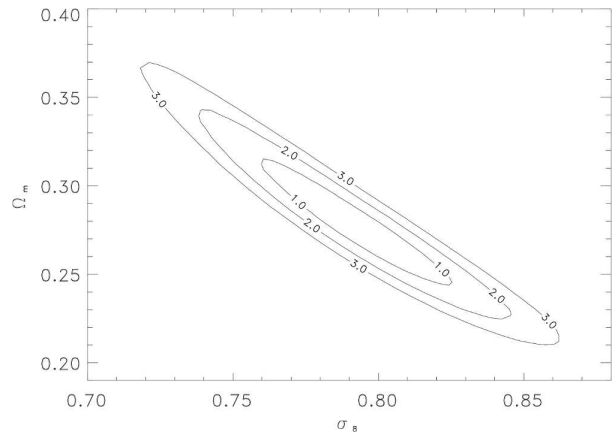}}
\end{center}
\caption{
{\bf (Upper panel)} X-ray luminosity function of the
\reflex\ I cluster survey. The luminosity function covers
not only the cluster regime but probes well into the 
parameter range of galaxy groups. Also shown is a model
prediction for a cosmology with the parameters, 
$\Omega_m=0.28$ and $\sigma_8=0.79$ assuming a flat
Universe and $H_0$=70~km/s/Mpc.
{\bf (Lower panel)} Cosmological constraints for the parameters
$\Omega_m$ and $\sigma_8$ for the above shown fitting of 
a model predicted X-ray luminosity function to the \reflex\ data.
The fit has only been performed for $L_X >0.25\times10^{44}$~erg/s,
because the calibration of the $L_X$-mass relation for the
group regime is too uncertain.
}
%the amplitude of the density fluctuation determined .}
\label{fig:2}
\end{figure}

The \reflex\ II galaxy cluster sample increases the size of these studies by a 
factor of two which already provides additional insights, e.g. into the way
the density contrast in the distribution of the clusters is amplified with 
respect to the density fluctuations of the matter, an effect called biasing.
Our results by Balaguera-Antolinez et al. (2011) based on \reflex\ II show that
this biasing increases with increasing X-ray luminosity of the clusters in
the sample exactly in the way predicted by the statistical theory of the large
scale structure. This result is a very important reassurance that our theory
of structure formation, on which the described cosmological tests rest, cannot
be grossly wrong. Further studies on the cosmic large-scale structure and
on the constraints of cosmological parameters with \reflex\ II are under way 
and they will pave the way for the data analysis to be performed on the much
larger \eros\ cluster sample.

\section{\eros\ cluster cosmology}

As it has been successfully shown with the \reflex\ catalogue,
a well-understood large catalogue of clusters of galaxies is 
an invaluable tool to understand the large-scale structure 
in our universe as well as cosmology.
The future X-ray mission, \eros\, aims to obtain
many more clusters by probing deeper flux and higher redshifts.
With the \eros\ telescopes we expect an improvement of 
two orders of magnitude in the sample size compared to the \rosat\
survey.

The relevant specification and expected performance of the \eros\ telescope 
can be briefly summarised as following. Its energy range covers
0.2-10 keV probing much wider energy range than {\it ROSAT}.
With its angular resolution of less than 13 arcsec on axis and
28 arcsec on average over the whole field of view, it improves
the sharpness of the X-ray imaging by a factor of 3 in 
linear scale.

\begin{figure}
\begin{center}
\resizebox{\hsize}{!}{\includegraphics{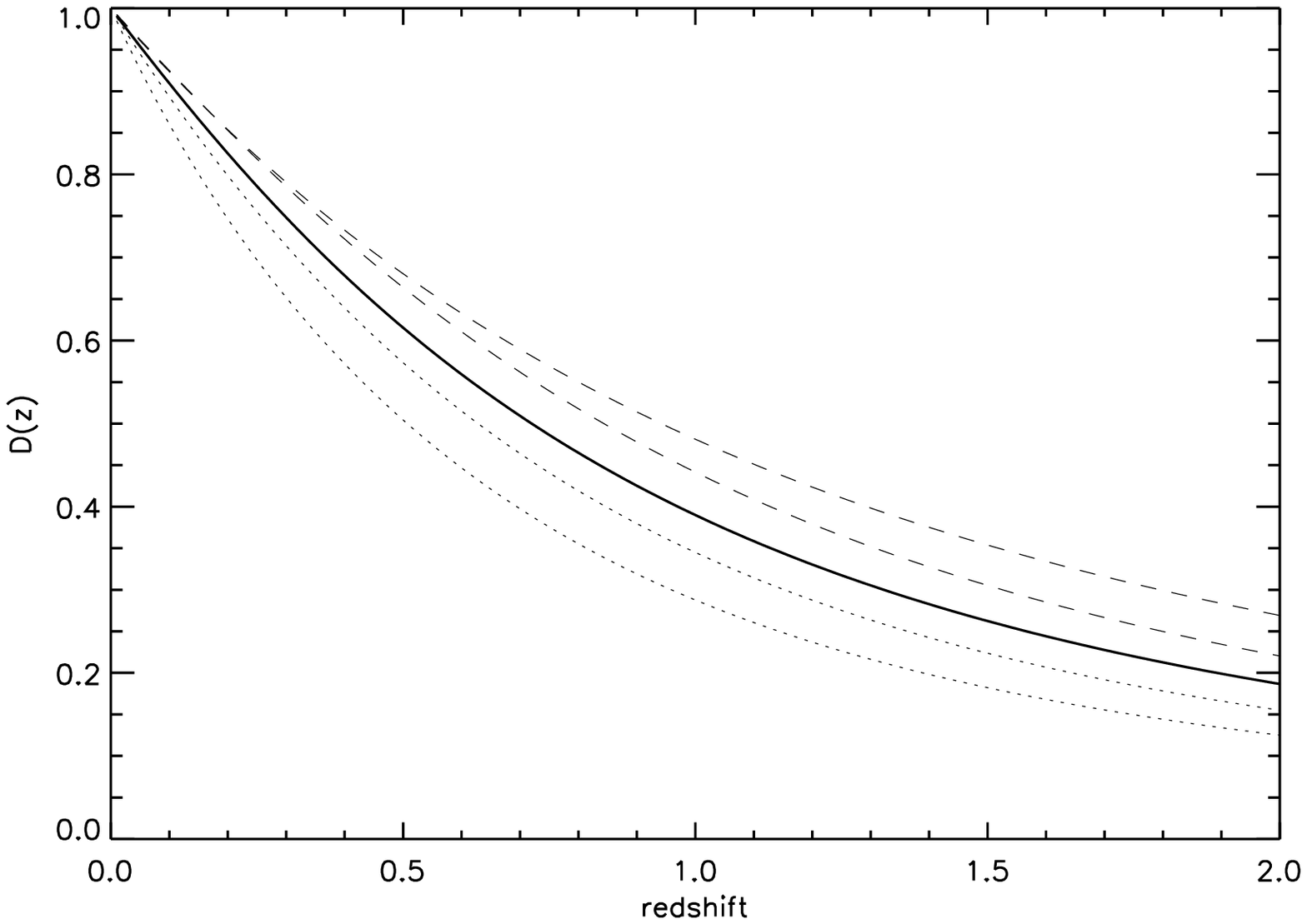}}
\resizebox{\hsize}{!}{\includegraphics{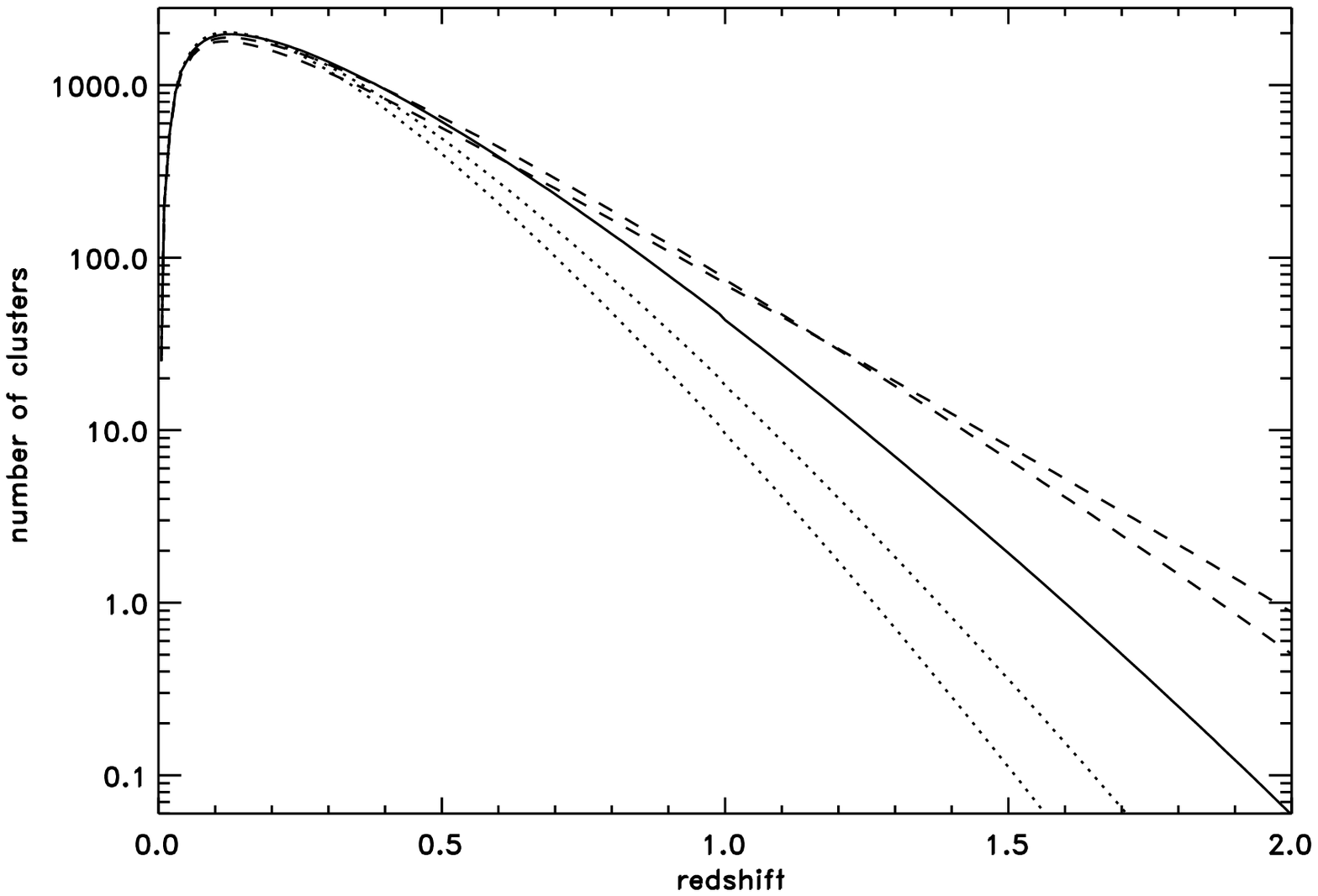}}
\resizebox{\hsize}{!}{\includegraphics{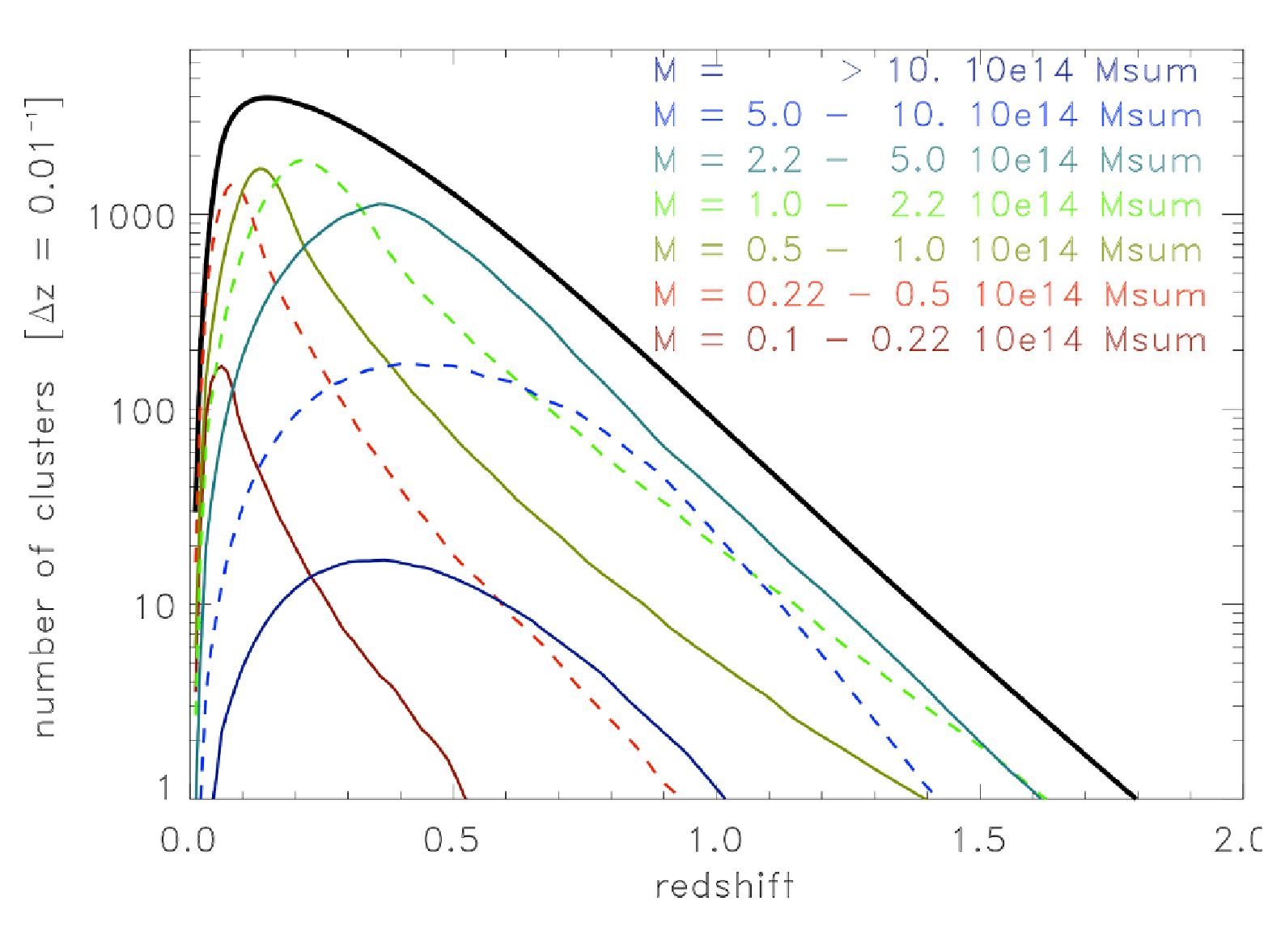}}
\end{center}
\caption{
{\bf (Upper panel)} Growth function as a function of redshift 
for various values of $w$. The solid line is for
the concordance model, i.e. $w$=-1, and the dashed lines above
are for $w$=-0.5 and -0.8, and the dotted lines below for $w$=-1.2, 
and -1.5.
{\bf (Middle panel)} Number count for varying $w$ values. The lines
of different values of $w$ correspond to those in the upper panel.
{\bf (Lower panel)} Number of clusters to be detected in the \eros\ survey
as a function of mass and redshift. One notes that galaxy 
groups are only detected in the nearby Universe while the most
massive clusters are no longer present at high redshift. 
The large number of clusters up to redshift 1 will allow 
a good discrimination between different Dark Energy models.}
\label{fig:3}
\end{figure}

With the wide redshift coverage of \eros\, 
clusters of galaxies also become sensitive probes of dark energy,
which is readily shown in Fig. 3. The upper panel shows
the growth function of the amplitude of dark matter fluctuations 
as a function of redshift. The solid line
represents the concordance model with the dark energy equation of
state parameter, $w$ = -1. 
Two curves above this model are for smaller values of $w$
and the two below are for the larger values. They differ by 25\%
incrementally. Apart from the lowest redshift it is quite clearly
shown that the growth function depends sensitively on the assumptions
about the dark energy parameter. The second panel in Fig. 3
shows our prediction for the detected number of clusters with
the \eros\ survey assuming that we can confirm the cluster
candidate with 100 photon counts. The curves correspond exactly
to the cosmological models shown in the upper panel of Fig. 3. 
By measuring the number count of clusters over a range of 
redshifts we can clearly distinguish different dark energy models, 
especially if higher redshifts are included. 

While there are some uncertainties in these predictions depending
on the exact cosmological model used and due to our currently
imperfect knowledge of the redshift evolution of the X-ray luminosity
-- mass scaling relations, we clearly note that one expects to detect
of the order of 100,000 clusters, with a good redshift coverage
up to unity and several hundred clusters at redshift up to $z\sim1.5$.

The types of clusters that the \eros\ survey is expected to see
is shown in the lower panel of Fig. 3. Here we show the detected
number of clusters as a function of redshift according to 
their masses. There are two clear effects shown. We see that
the most massive clusters are becoming rare at the highest redshifts.
This is due to the fact that clusters of this mass are just starting
to form. The other point shown by the clusters with
the least mass is that we will be able to detect many local
groups, however, given the sensitivity of the instrument, 
we run out of them soon after redshift, $z=0.2$.

\section{\eros\ cluster astrophysics}

The \eros\ survey will also allow us to greatly improve our capabilities of
astrophysical studies of clusters. The basis of many of such studies will be 
the fact that we will detect several thousand clusters with more than 1000
and more than 10000 clusters with more than 500 source photon counts. For 
these clusters more detailed information can be obtained, for example, 
on the temperature of the intracluster medium through a spectral analysis 
of the X-ray radiation as shown in Fig. 4. 
The simulation shown in the figure was performed for a cluster with
a temperature of 4 keV at $z=0.2$ for 1000 source counts in a 2 ks
exposure. A realistic instrumental and sky background spectrum were 
added and subsequently subtracted with a different photon statistical
realisation.

With increasing temperature the uncertainty of the temperature measurement
gets larger due to the fact that at lower temperature there are more
features in the spectrum which change with temperature variations.

The X-ray determined cluster 
temperature is a much better mass proxy than the X-ray luminosity 
(e.g. Kravtsov et al. 2006). The \eros\ survey will thus allow very detailed 
studies of e.g. the X-ray luminosity - temperature relation with an assessment
of several important bias effects, which will also lead to an improvement 
of the mass calibration for the cosmological modeling described above. 
These photon numbers will also allow important statistical studies on the 
morphology of galaxy clusters, through the study of substructure 
characterizations as shown in  B\"ohringer et al. (2011). Such studies will
in the end help to characterize the degree of virial relaxation of the clusters
as a function of their mass and their redshift, an information that will give
insights into the formation history of galaxy clusters and further improve our
understanding of making cosmological predictions of the observable statistics 
of the galaxy cluster population for cosmological tests. Many more 
interesting studies will be possible, which cannot be described in this 
short article.

\begin{figure}
\begin{center}
\resizebox{\hsize}{!}{\includegraphics[height=4cm]{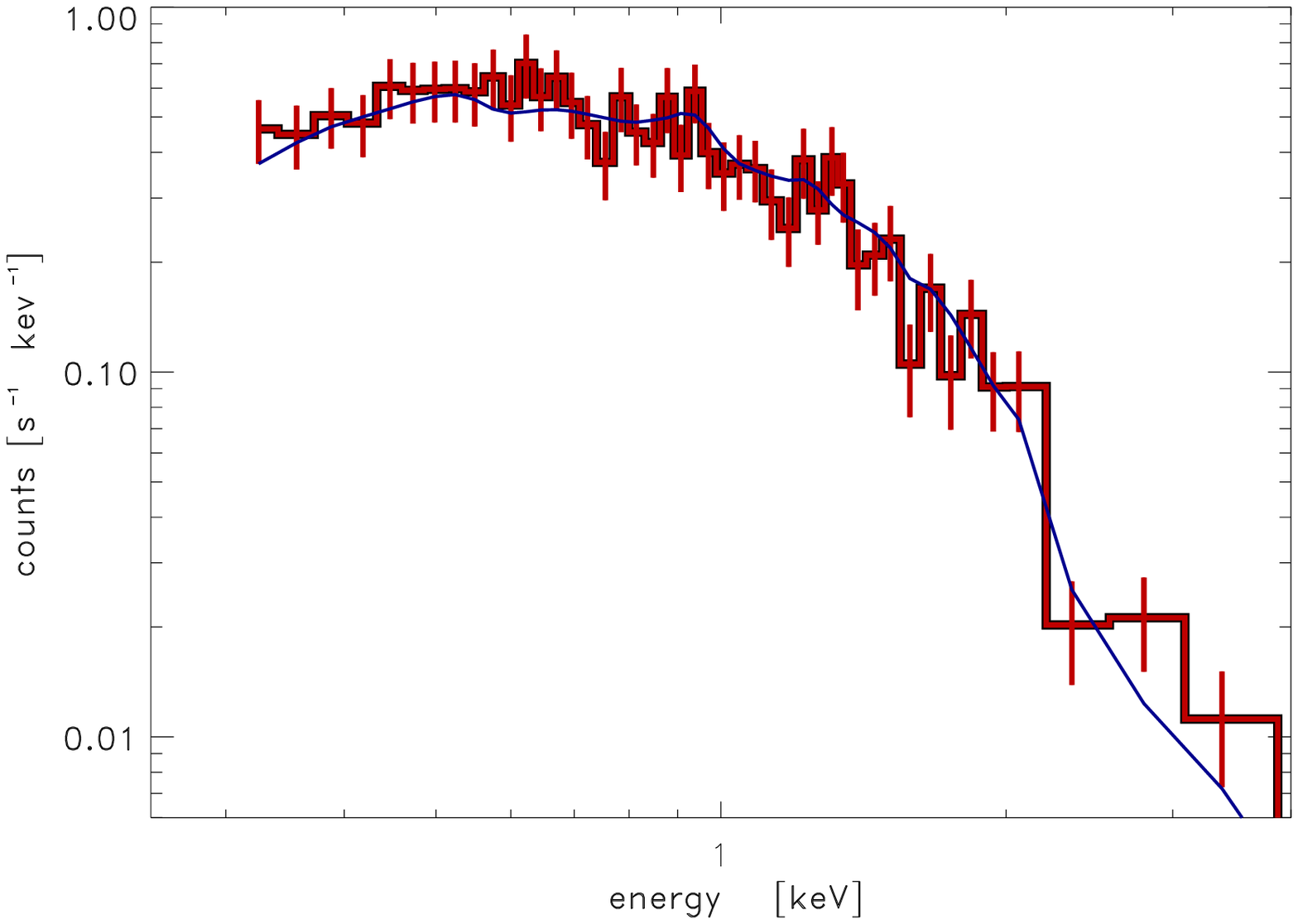}}
\resizebox{\hsize}{!}{\includegraphics[height=4cm]{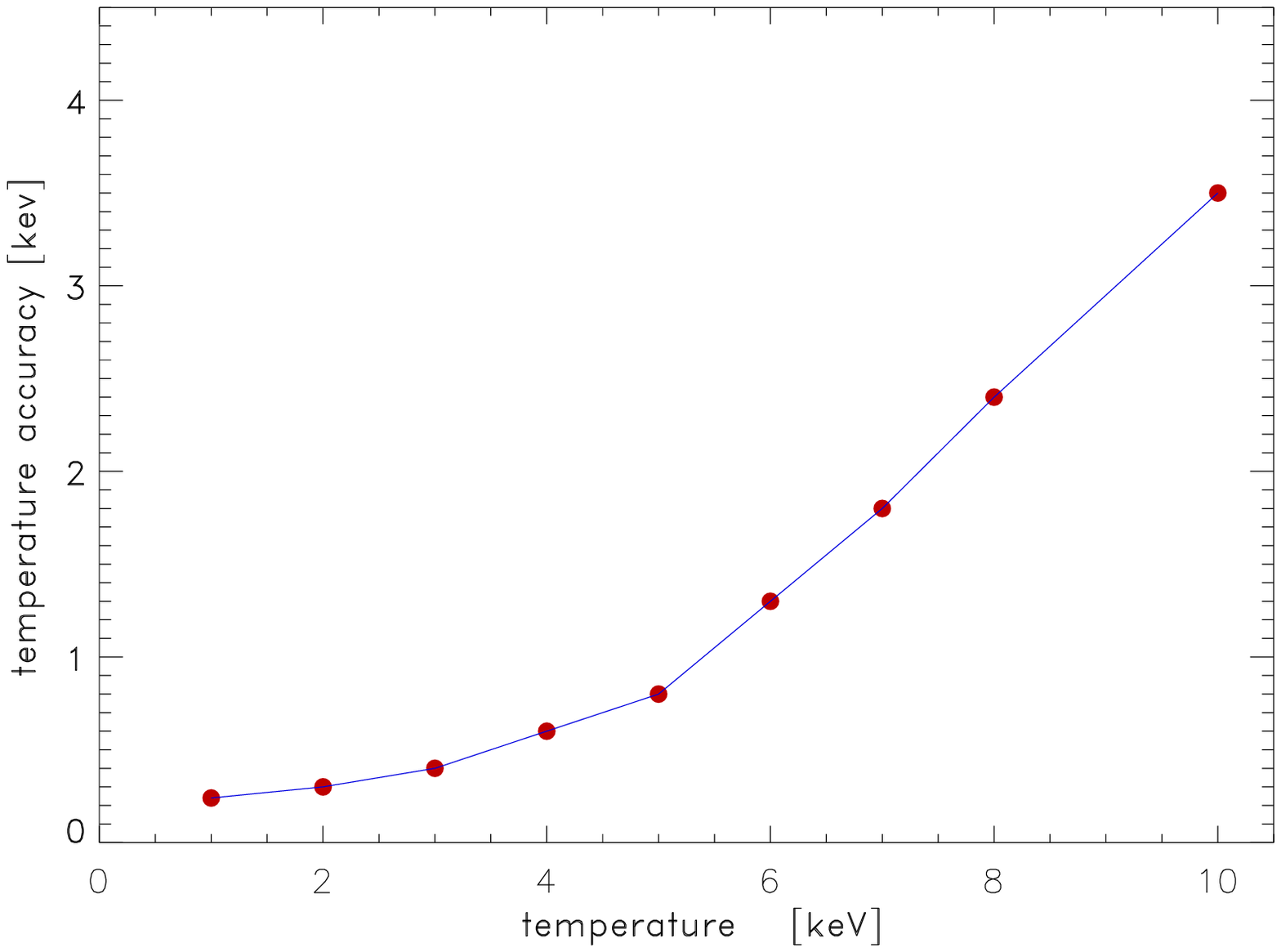}}
\end{center}
\caption{
{\bf (Upper panel)} Simulated \eros\ spectrum of a cluster with an
ICM temperature of 4 keV at $z=0.2$ with a metal abundance of 0.3. 
For the simulation we assumed that
the cluster was detected with 1000 source counts with an \eros\ survey
exposure of 2 ks using an extraction radius of 3 arcmin. The expected
X-ray background was included in the simulations and a background 
model was subtracted from the data shown in the plot. 
{\bf (Lower panel)} The accuracy, in one $\sigma$, with which the 
temperature can be recovered from a detection of 1000 source counts. 
Spectra of the kind
shown in the upper panel with 1000 source counts with different input
temperature are used to determine the temperature and its uncertainty
with the XSPEC software.
}
\label{fig:4}
\end{figure}

\section{Summary}

X-ray clusters of galaxies with a well-understood selection function
have been one of the most successful tools to investigate variety of 
astrophysical and cosmological questions. With the expected launch 
of the \eros\ instrument in 2014 we anticipate a leap forward in 
understanding the evolution of the large-scale structure and dark energy.
We will also obtain a comprehensive picture about the structural
evolution of clusters from $z=1$ to the present. Pointed observations
with \eros\ that will be possible after the completion of the survey 
will enable deeper studies of well-selected study samples of
clusters from the \eros\ survey.

\acknowledgements
We thank the referee and the organising committees for this 
conference. GC acknowledges the support from Deutsches Zentrum f\"ur 
Luft und Raumfahrt (DLR). We acknowledge support from the DfG 
Transregio Program TR33 and the Munich Excellence Cluster 
``Structure and Evolution of the Universe''.

%\newpage%%%%%%%%%%%%%%%%%%%%%%%%%%%%%%%%%%%%%%%%%%%%%%%%%%%%%%

\end{document}